\documentstyle[12pt]{article}
\catcode`\@=11
\renewcommand{\thefootnote}{\fnsymbol{footnote}}

\begin{document}

\begin{titlepage}

\hfill{DFPD/95/TH/38}

\hfill{hep-th/9506181}

\vspace{0.3cm}

{\centerline{\large \bf KOEBE 1/4-THEOREM AND INEQUALITIES}}

\vspace{0.8cm}

{\centerline{{\large \bf  IN N=2 
SUPER-QCD}\footnote[2]{Partly
supported by the European Community Research
Programme {\it Gauge Theories, applied supersymmetry and quantum 
gravity}, contract SC1-CT92-0789}}}

\vspace{1.5cm}

{\centerline{\sc MARCO MATONE}}

\vspace{1cm}

\centerline{\it Department of Physics ``G. Galilei'' - Istituto Nazionale di 
Fisica Nucleare}
\centerline{\it University of Padova}
\centerline{\it Via Marzolo, 8 - 35131 Padova, Italy}
\centerline{matone@padova.infn.it}

\vspace{2cm}

\centerline{\sc ABSTRACT}

\vspace{1cm}

The critical curve ${\cal C}$ on which ${\rm Im}\,\hat\tau =0$,
 $\hat\tau=a_D/a$, determines 
hyperbolic domains whose Poincar\'e metric is constructed in terms of
$a_D$ and $a$. 
We describe ${\cal C}$ 
in a parametric form related to a Schwarzian equation
and prove new relations for $N=2$ Super $SU(2)$ Yang-Mills.
In particular, using
the Koebe 1/4-theorem and Schwarz's lemma,
we obtain inequalities involving $u$, $a_D$ and $a$,
which seem related to the 
Renormalization Group.
Furthermore, we obtain a closed form for the prepotential
as function of $a$.
Finally, we show that $\partial_{\hat\tau}
\langle {\rm tr}\,\phi^2\rangle_{\hat \tau}
={1\over 8\pi i b_1}\langle \phi\rangle_{\hat\tau}^2$,
where $b_1$ is the one-loop coefficient of the beta function.

\end{titlepage}

\newpage

\setcounter{footnote}{0}

\renewcommand{\thefootnote}{\arabic{footnote}}

\noindent
{\bf 1.} The effective action of the
low-energy limit of $N=2$ super Yang-Mills, solved exactly in 
\cite{SW1}, is described in terms of the prepotential ${\cal F}$ \cite{S}
\begin{equation}
S_{eff}={1\over 4\pi}{\rm Im}\,\left(\int d^2\theta d^2\bar\theta
\Phi^i_D\overline \Phi_i+{1\over 2}\int d^2\theta
\tau^{ij}W_iW_j\right),
\label{01}\end{equation}
where $\Phi^i_D\equiv {\partial {\cal F}/\partial \Phi_i}$ and
$\tau^{ij}\equiv {\partial^2 {\cal F}/\partial \Phi_i \partial \Phi_j }$.
Let us denote by $a_i\equiv \langle \phi^i\rangle$ and
$a^i_D\equiv \langle \phi_D^i\rangle$ 
the vevs of the scalar component of the chiral 
superfield.
For gauge group $SU(2)$ the moduli space of quantum vacua, parameterized by 
$u\equiv\langle{\rm tr}\, \phi^2\rangle$, is 
$\Sigma_3={\bf C}\backslash\{-\Lambda^2,\Lambda^2\}$, 
the Riemann sphere 
$\widehat {\bf C}={\bf C}\cup \{\infty\}$ with punctures at 
$\pm \Lambda^2$ and 
$\infty$, where $\Lambda$ is the dynamically generated scale.
 It turns out 
that \cite{SW1} (we set $\Lambda=1$)
\begin{equation}
a_D=\partial_a{\cal F}=
{\sqrt 2\over \pi}\int_1^u {dx \sqrt{x-u}\over \sqrt{x^2-1}},
\qquad a={\sqrt 2\over \pi}\int_{-1}^1 {dx \sqrt{x-u}\over 
\sqrt{x^2-1}}.
\label{14}\end{equation}
A crucial property of $a_D$ and $a$ is that they satisfy the equation
\cite{KLT} (see also \cite{mm})
\begin{equation}
\left[4(u^2-1)\partial_u^2+1\right]a_D=0=
\left[4(u^2-1)\partial_u^2+1\right]a.
\label{2}\end{equation}
This equation is the ``reduction'' of the uniformizing equation 
for $\Sigma_3$ \cite{CerDaFe}\cite{KLT}\cite{mm}
\begin{equation}
\left[4(1-u^2)^2\partial_u^2 +u^2+3 \right]\psi=0,
\label{11bis}\end{equation}
which is satisfied
by $\sqrt{1-u^2}\partial_ua_D$ and $\sqrt{1-u^2}\partial_ua$.

Let us summarize the main results in \cite{mm}. First of all
it has been shown that 
\begin{equation}
u=\pi i ({\cal F}-a\partial_a{\cal F}/2),
\label{81}\end{equation}
that is
\begin{equation}
{\cal F}\left(\langle \phi\rangle\right)=
{1\over \pi i} \langle{\rm tr}\, \phi^2\rangle+
{1\over 2}\langle \phi\rangle \langle \phi_D\rangle.
\label{fst}\end{equation}
In order to specify 
the functional dependence of $u$
 we set $u={\cal G}_1(a)$,
$u={\cal G}_2(\hat \tau)$ and $u={\cal G}_3(\tau)$
where $\hat\tau=a_D/a$ and $\tau=\partial_a^2 {\cal F}$.
By Eq.(\ref{2}) we have 
\begin{equation}
(1-{\cal G}_1^2)\partial_a^2{\cal G}_1+{a\over 4}
\left(\partial_a{\cal G}_1\right)^3=0,
\label{1}\end{equation}
that by (\ref{81}) 
provides recursion relations for the instanton contribution
and implies
\begin{equation}
\partial_a^3 {\cal F}=
{\pi^2\left(a \partial_a^2{\cal F}-\partial_a{\cal F}\right)^3
\over 16\left[1+\pi^2 \left({\cal F}- 
{a}\partial_a{\cal F}/2\right)^2\right]}.
\label{gdtf}\end{equation}
By (\ref{14}) we have $a(u=-1)= -i4/\pi$ and  $a(u=1)=4/\pi$ so that
the initial conditions for the 
second-order equation (\ref{1}) are
${\cal G}_1(-i4/\pi)=-1$ and ${\cal G}_1(4/\pi)=1$.

In this paper we will investigate some consequences of the relation
(\ref{81}).
This relation allows us to find the 
differential equation satisfied by the functions ${\cal G}_k$
and implies a new relation which involves theta-functions and 
the prepotential ${\cal F}$.
Furthermore, we investigate the structure
of ${\cal F}$ as function of $a$ that, 
although its explicit expression is still unknown, we give in a
closed form in Eq.(\ref{aaiqwnd}). 

In our investigation the Seiberg-Witten critical curve ${\cal C}$,
on which ${\rm Im}\,\hat\tau=0$, plays a crucial role.
This curve determines hyperbolic 
domains. 
In particular, using uniformization theory, we will construct the natural 
(Poincar\'e) metrics on the quantum moduli space and on the
hyperbolic domain inside ${\cal C}$.
Such metrics should be also useful in 
finding the building blocks for the non-holomorphic part
 (higher order derivatives) of
the effective action.
General theorems
concerning univalent functions, such as Schwarz's lemma and 
the Koebe 1/4-theorem, imply inequalities that for the case at hand take
a simple form in terms of the vevs of $\phi$, $\phi_D$ and ${\rm tr}\,
\phi^2$.
We will suggest that such inequalities
are related to the Renormalization 
Group and will discuss a possible connection with the uncertainty principle.
In obtaining such inequalities we introduce the Euclidean distance
between points in the quantum moduli spaces. The structure of this 
distance is investigated making use of
Eq.(\ref{81}) which also implies the relation
$\partial_{\hat\tau}
\langle {\rm tr}\,\phi^2\rangle_{\hat \tau}
={1\over 8\pi i b_1}\langle \phi\rangle_{\hat\tau}^2,$
where $b_1$ is the one-loop coefficient of the beta function.

We note that some properties
of the critical curve have been
investigated in \cite{SW1}\cite{F}\cite{AFS} whereas related physical
aspects have been discussed in  \cite{SW1}\cite{LindstromRocek}.

\vspace{0.6cm}

\noindent
{\bf 2.} The critical curve ${\cal C}$ can be seen as the curve
on which the torus with modular parameter $\hat \tau$ degenerates.
Let us consider the mass of a dyon hypermultiplet
\begin{equation}
M_{n_mn_e}=\sqrt 2 |n_ma_D+n_ea|=\sqrt 2|a||n_m\hat\tau+n_e|,
\label{09}\end{equation}
where $n_e$ and $n_m$ are the electric and magnetic charges 
respectively.
$M_{n_mn_e}^2$ is related to the eigenvalues of the Laplacian on the 
$\hat\tau$-torus. More precisely we have
the Schr\"odinger equation
\begin{equation}
\Delta\psi_{n_m n_e}=E_{n_m n_e}\psi_{n_m n_e},
\label{gdhgf}\end{equation}
where $\Delta=-2\partial_{\bar z}\partial_z$ is the Laplacian on the 
torus and $\psi_{n_m n_e}=\sqrt 2\cos 2\pi(n_mx-n_ey)$,
$z=x+\hat\tau y$. One has 
$E_{n_mn_e}=2\pi^2|n_m\hat \tau +n_e|^2/
({\rm Im}\, \hat\tau)^2$, so that
\begin{equation}
E_{n_m n_e}={\pi^2 M_{n_mn_e}^2\over |a|^2({\rm Im}\, \hat\tau)^2}.
\label{energy}\end{equation}
Notice that  (\ref{gdhgf}) admits the following interpretation.
One can consider the theory $N=1$ in $D=6$. Compactifying two 
dimensions on the torus $\hat \tau$, one has
$Z\sim P_5+iP_6$ so that in the massless sector $P^2=0$
\begin{equation}
\Box_6\psi=0\Longrightarrow \Box_4\psi=\Delta_2\psi=|Z|^2\psi,
\qquad \qquad \Delta_2={2\pi^2\over |a|^2({\rm Im}\, \hat \tau)^2}\Delta
\label{gdftaaa}\end{equation}

In crossing the curve ${\cal C}$ 
a BPS-saturated particle of given charges 
can appear or disappear. Eqs.(\ref{gdhgf})(\ref{energy}) 
show that the tori $\hat\tau\in {\cal C}$ correspond to  critical 
points for the structure of the energy eigenvalues.
In \cite{F}\cite{AFS} it has been
shown that inside ${\cal C}$, that is in the
domain $A$ containing the point $u=0$ and such that ${\cal C}=\partial A$,
 one has ${\rm Im}\,\hat\tau<0$. We will show that 
${\cal C}$ can be parametrically described by the solution of a Schwarzian 
equation. Eventually we will obtain inequalities which resemble a sort of 
uncertainty relations for QFT and where the critical curve plays a crucial 
role. In order to find the differential equation associated to ${\cal C}$ we 
first recall that $\tau$ corresponds to the inverse of 
the map which uniformizes $\Sigma_3$ \cite{mm}. An important point is that 
both $\tau$ and $\hat \tau$ have $\Gamma(2)$-monodromy implying that the 
structure of the associated fundamental domains $D_1$ and $D_2$
differ for the value of the 
opening angle at the cusps \cite{AFS} ($0$ and $2\pi$ respectively). 
To describe the critical curve, we first note that by 
definition (see also \cite{AFS})
\begin{equation}
{\cal C}=\left\{u={\cal G}_2(\hat \tau)|\hat \tau\in [-1,1]\right\}.
\label{04}\end{equation}
On the other hand, the dependence of $u$ on $\hat \tau$ can be 
determined by solving 
a differential equation which follows from (\ref{2}). 
Using the following property of the 
Schwarzian derivative
\begin{equation}
\{y(x),x\}={y'''\over y'}-{3\over 2}
\left({y''\over y'}\right)^2=-{y'}^2\{x(y),y\},
\label{oioj}\end{equation}
one has 
$\left\{{\cal G}_2,\hat \tau\right\}=-
(\partial_{\hat \tau}{\cal G}_2)^2\left\{\hat \tau,{\cal G}_2\right\}$.
To obtain the differential equation satisfied by ${\cal G}_2$
we recall that if $\psi_1,\psi_2$ are linearly independent 
solutions of 
$\partial_x^2\psi(x)+P(x)\psi(x)=0$, then $\{\psi_1/\psi_2,x\}=2P$,
which follows
by $f^{1/2}\partial_x f^{-1}\partial_x 
f^{1/2}=\psi_k^{-1}\partial_x\psi_k^2\partial_x\psi_k^{-1}
=\partial_x^2-\psi_k''/\psi_k=\partial_x^2+P$,
$k=1,2$, $f=\psi_1/\psi_2$ (note that 
$\psi_2(x)=A\psi_1(x)+B\psi_1(x)\int^x\psi_1^{-2}$, $B\ne 0$).
Thus, since $\hat \tau$ is the ratio of two solutions of 
(\ref{2}), we have
\begin{equation}
2(1-{\cal G}_2^2)\left\{{\cal G}_2,\hat \tau\right\}=
\left(\partial_{\hat \tau}{\cal G}_2\right)^2.
\label{4}\end{equation}
An explicit computation gives $\hat\tau(u=-1)=\pm 1$, 
$\hat \tau (u=1)=0$ (observe that $-1$ and $+1$ are identified after 
factorizing the image of the $\hat \tau$-map by $\Gamma(2)$), so that
\begin{equation}
{\cal G}_2(-1)=
{\cal G}_2(1)=-1,\qquad
{\cal G}_2(0)=1.
\label{odiq}\end{equation}
Therefore the critical curve is given by 
(\ref{04}) with ${\cal G}_2$ solution of (\ref{4}) and initial 
conditions (\ref{odiq}). 
The solutions of Eqs.(\ref{1})(\ref{4}) 
should be related to the $\wp$-function. To show
this we use again (\ref{oioj}) so that 
$\left\{{\cal G}_3,\tau\right\}=-
\left({\partial_{\tau}}{\cal G}_3\right)^2\left\{\tau,{\cal G}_3\right\}$
and by (\ref{11bis})
\begin{equation}
2(1-{\cal G}_3^2)^2\left\{{\cal G}_3,\tau\right\}=
-\left(3+{\cal G}_3^2\right)\left(\partial_\tau{\cal G}_3\right)^2.
\label{11tris}\end{equation}
The fact that $\tau$ and $\hat \tau$ have the same monodromy
and Eq.(\ref{odiq}) imply that
(see also \cite{AFS}) 
\begin{equation}
{\cal G}_3(-1)=
{\cal G}_3(1)=-1,\qquad
{\cal G}_3(0)=1,
\label{odiqbis}\end{equation}
which also follows by an explicit computation (use (\ref{14}) and recall
that $\tau=a_D'/a'$).
A way to find the solution of (\ref{11tris})
with initial conditions (\ref{odiqbis}) is to consider
$u$ as the 
uniformizing map. In the case of
$\Sigma_3$ we have \cite{Ford}
\begin{equation}
u={\cal G}_3(\tau)={2\wp\left({1+\tau\over 2}\right)-
\wp\left({\tau\over 2}\right)
- \wp\left({1\over 2}\right)\over \wp\left({\tau\over 
2}\right)-
\wp\left({1\over 2}\right)}=1- 2\left[\Theta_2(0|\tau)\over 
\Theta_3(0|\tau)\right]^4,
\label{saouihkj}\end{equation}
that by the ``inversion formula'' 
(\ref{81}) implies the new relation
\begin{equation}
\pi i \left({\cal F}-{a\over 2}\partial_a {\cal F}\right)=
1- 2\left[\Theta_2\left(0|\partial_a^2{\cal F}\right)\over 
\Theta_3\left(0|\partial_a^2{\cal F}\right)\right]^4,
\label{saouihkjwe}\end{equation}
showing that such a combination of theta-functions acts on 
$\partial^2_a{\cal F}$ as integral operators.

By (\ref{saouihkj}) the problem of finding the explicit solutions of 
Eqs.(\ref{1})(\ref{4}) is equivalent to the problem of finding the 
explicit dependence of $\tau$ as function of $a$ and $\hat\tau$ 
respectively. In this context we note that once Eq.(\ref{1}) is solved, 
we can use Eq.(\ref{81}) to obtain 
the explicit dependence of ${\cal F}$ on $a$, namely 
\begin{equation}
{\cal F}(a)={2i\over \pi}a^2\int^a_{a_0}dx {\cal G}_1(x)x^{-3}
-{i u_0\over \pi a_0^2}a^2+{a_{D0}\over 2a_0}a^2,
\label{iqwnd}\end{equation}
where $u_0$ is an arbitrary point on the compactified moduli space
$\overline \Sigma_3=\widehat{\bf 
C}$,
and $a_0\equiv a(u_0)$, $a_{D0}\equiv a_D(u_0)$. Choosing $u_0=1$ we 
have
\begin{equation}
{\cal F}(a)={2i\over \pi}a^2\int^a_{4/\pi}dx {\cal G}_1(x)x^{-3}
-{i \pi \over 16} a^2.
\label{aaiqwnd}\end{equation}

\vspace{0.6cm}

\noindent
{\bf 3.} In \cite{SW1} it has been emphasized that the properties of the 
metric\footnote{We will show that $e^\varphi$
is the Poincar\'e metric on $\Sigma_3$ so that
$e^{-\varphi/2}$
is a ``non-chiral'' solution of the 
uniformizing equation (\ref{11bis}) (see \cite{M}).}
\begin{equation}
ds^2={\rm Im}\, \left({\partial^2 {\cal F}\over \partial 
a^2}\right)|da|^2= {e^{-\varphi/2}\over 2\pi|1-u^2|}|du|^2,
\label{15}\end{equation}
are at heart of the physics. Actually, the natural framework to 
investigate these properties is uniformization theory. An
interesting aspect of the results in \cite{SW1} is that 
the classical moduli space of the theory
is the Riemann sphere with a puncture whereas in the quantum case one 
has the Riemann sphere with three punctures. Thus, since by Gauss-Bonnet 
formula for $n$-punctured Riemann spheres one has 
$\int_{\Sigma_n}\sqrt g R_g=2\pi(2-n)$, there is a 
``transition'' from positively (classical moduli) to negatively 
(quantum moduli) curved spaces. This transition 
makes it evident that quantum aspects are related to deep
aspects concerning uniformization theory. In particular, one can apply basic
inequalities, such as the Koebe 1/4-theorem and Schwarz's lemma, 
which are at heart of the
theory of univalent functions (i.e. uniformization, Teichm\"uller spaces 
etc.). 

We now use the prepotential ${\cal F}$ to construct 
the positive definite  metric
\begin{equation}
ds_P^2={\left|{\partial^3{\cal F}/\partial a^3}\right|^2\over
\left({\rm Im}\, {\partial^2{\cal F}/ \partial a^2}\right)^2}
|da|^2={\left|{\partial^3{\cal F}/\partial u\partial a \partial a}
\right|^2\over
\left({\rm Im}\, {\partial^2{\cal F}/ \partial a^2}\right)^2}
|du|^2=e^\varphi |du|^2.
\label{16bb}\end{equation}
Let $H=\{w|{\rm Im}\, w>0\}$ be the upper half plane endowed with the 
Poincar\'e metric $ds^2_P=({\rm Im}\, w)^{-2}|dw|^2$. Since $\tau$ is 
the inverse of the uniformizing map $J_H: H\to \Sigma_3$,
it follows that
$e^\varphi$ is the Poincar\'e metric on $\Sigma_3$ so that
$\varphi$ satisfies the Liouville equation $\varphi_{u\bar u}
=e^{\varphi}/2$. 

To prove the last equality in (\ref{15}) observe that 
$\partial_u\partial_a^2 {\cal F}=(a'a_D''-a_Da'')/{a'}^2$
where $'\equiv\partial_u$. By (\ref{2})
and using \cite{mm} 
\begin{equation}
aa'_D-a_Da'={2i\over\pi},
\label{ohaio}\end{equation}
it follows that
\begin{equation}
{\partial^3{\cal F}\over \partial u\partial a\partial a }
= {1\over 2\pi i {a'}^2(1-u^2)},
\label{idhpo}\end{equation}
which is equivalent to (\ref{gdtf}). The
last equality in (\ref{15}), that is
$e^{-\varphi/2}= 2\pi |a'|^2 |1-u^2|{\rm Im}\, (\partial_a^2{\cal F})$,
follows by the definition of $e^{\varphi}$ given in (\ref{16bb}).

We now observe that the $A$-domain can be seen as a distortion 
by a regular (note that $\infty\notin A$) univalent function
of the Poincar\'e disk $\Delta=\{z||z|<1\}$.
This remark suggests to apply 
distortion theorems for hyperbolic domains.
A basic point in our construction is 
that the Poincar\'e metric
on $A$ is easily identified in terms of the vevs of $\phi$ and $\phi_D$.
In particular, it can be explicitly expressed 
in terms of the function $\hat \tau(u)$ which maps $\Sigma_3$ to $D_2$.

Setting $w=\hat\tau(u)$ in 
$ds^2_P=({\rm Im}\, w)^{-2}|dw|^2$ (which is divergent for $w\in {\bf R}$),
we obtain the Poincar\'e metric on the hyperbolic domain $A$ 
\begin{equation}
ds^2_P=-4{|aa'_D-a_Da'|^2\over (a_D\bar a -\bar a_D a)^2}|du|^2=
-{16\over \pi^2}
{1\over (a_D\bar a -\bar a_D a)^2}|du|^2=e^{\varphi_A}|du|^2,
\label{khjui}\end{equation}
where we used (\ref{ohaio}).
We now apply the general construction for hyperbolic domains
(see for example \cite{NagLehto}) to show 
that\footnote{We stress that similar results hold in
the complementary 
domain $\widehat {\bf C}\backslash A$.}
\begin{equation}
1\le e^{\varphi_A(u,\bar u)} (\Delta_A u)^2\le 4,
\label{uncertaintyrelations}\end{equation}
where $\Delta_Au$ denotes the Euclidean distance from 
$u\in A$ to the critical curve ${\cal C}=\partial A$.

We note that a similar geometrical uncertainty relation appears in the 
description of the cutoff $(\Delta z_{min})^2$ in 2D quantum gravity \cite{M}.
In particular, it has been shown that in a hyperbolic domain $D$ one has
\begin{equation}
(\Delta z)^2\ge {\epsilon \over 4}e^{-\sigma_D} (\Delta_D z)^2,
\label{idqlk}\end{equation}
where $\epsilon=(\Delta s_{min})^2=e^{\varphi_D+\sigma_D}
(\Delta z_{min})^2$ is the minimal invariant length, $e^{\varphi_D}$ is the 
Poincar\'e metric on $D$ and $\sigma_D$ is the 
Liouville field. Eq.(\ref{idqlk}) provides the relation between the 
minimal length in configuration space $z$, the structure of the
boundary $\partial D$ and the Liouville field.

To prove (\ref{uncertaintyrelations}) we need Schwarz's lemma and the 
Koebe 1/4-theorem (see for example \cite{Nehari1}).
Let $f(z)$ be an
analytic and regular function in $\Delta$ 
vanishing in zero. Schwarz's lemma 
states that if $|f(z)|\le 1$,  $z\in \Delta$, then $|f(z)/z|\le 1$, 
$z\in\Delta$, where equality can hold only if $f$ and $z$ differ by a 
phase.
Setting $z=0$ in the expansion $f(z)=f'(0)z+f''(0)z^2/2!+\ldots$, we obtain
\begin{equation}
|f'(0)|\le 1.
\label{xiwx}\end{equation}
Let $u$ be a point in $A$ and $F$ a conformal map of $A$ onto $\Delta$.
Since the Poincar\'e metric on $\Delta$ is $ds^2_P=4(1-|z|^2)^{-2}|dz|^2$,
we have
\begin{equation}
e^{\varphi_A(u,\bar u)}=4|F'(u)|^2.
\label{d092}\end{equation}
Setting $f(z)=F\left(u+z\Delta_Au\right)$, 
by (\ref{xiwx}) we have $|F'(u)|\Delta_Au\le 1$, so that
(\ref{d092}) implies 
$e^{\varphi_A(u,\bar u)} (\Delta_A u)^2\le 4$.
The other inequality in (\ref{uncertaintyrelations}) is an application
of the Koebe 1/4-theorem, 
a consequence of the area theorem. It states that if $f$ is a regular and
univalent function in $\Delta$ with normalization $f(0)=0$, $f'(0)=1$ 
(functions with such properties constitute the ${\cal S}$-class), 
then $f(\Delta)$
contains the open disk of radius $1/4$ center $0$.
Let $g$ be a conformal mapping of $\Delta$ onto $A$ with $g(0)=u$,
we have
\begin{equation}
e^{\varphi_A(u,\bar u)}=4|g'(0)|^{-2}.
\label{paraponziponzipero}\end{equation}
We now set $G(z)= {(g(z)-g(0))/g'(0)}$, so that 
$G\in {\cal S}$,
and observe
that since by construction
$g(\Delta)$ contains the open disk of radius $\Delta_Au$ center $u$,
it follows that $G(\Delta)$ 
 contains the open disk of radius 
$\Delta_Au/|g'(0)|$ center $0$.
On the other hand, by the Koebe 
$1/4$-theorem, $G(\Delta)$ contains the open disk of 
radius $1/4$ with center $0$, so that $\Delta_Au/|g'(0)|\ge 1/4$
and (\ref{paraponziponzipero}) implies
$e^{\varphi_A(u,\bar u)} (\Delta_A u)^2\ge 1$.

\vspace{0.6cm}

\noindent
{\bf 4.} Up to now we considered unit where $\hbar=1$. However, 
the structure of (\ref{uncertaintyrelations}) suggests 
performing a dimensional 
analysis setting the Euclidean distance dimensionless.
Since ${\cal F}$ has the dimensions of $\hbar$ and being
$a_D=\partial_a{\cal F}$, by (\ref{uncertaintyrelations}) we have
\begin{equation}
\hbar \Delta_A\langle{\rm tr}\,\phi^2\rangle
\leq 
{\pi}{\rm Im}\,\left(\langle\phi\rangle
\overline{\langle\phi_D\rangle}\right)\leq
2\hbar \Delta_A\langle{\rm tr}\,\phi^2\rangle,
\label{uncertainty1}\end{equation}
where we used the fact that ${\rm Im}\,a_D/a<0$.
Let us denote by ${\cal C}_{d}$ the curve in $A$ on which
$\Delta_A\langle{\rm tr}\,\phi^2\rangle=d$. On 
${\cal C}_{1/2}$ Eq.(\ref{uncertainty1}) has the structure
\vspace{0.2cm}
\begin{equation}
{\hbar\over 2}\le \pi{\rm Im}\,\left(\langle\phi\rangle
\overline{\langle\phi_D\rangle}\right)\le \hbar.
\label{ANTONIAALICEROCCO}\end{equation}

In order to investigate the physical meaning of (\ref{uncertainty1})
we should first discuss two aspects. The first one
concerns the structure of the Euclidean distance
$\Delta_Au$.
Let us denote by $v$ the points in ${\cal C}$
and by $x$ the values
of $\hat\tau$ such that $v={\cal G}_2(x)$, so that by (\ref{04})
${\rm Im}\, x=0$ and $x\in [-1,1]$.
Note that by definition 
$\Delta_Au=|u-v_0|$, where $u\in A$ and $v_0$ is the minimum of $|u-v|$.
In order to determine $v_0$ one should first solve
the equation 
\begin{equation}
\partial_x \left|{\cal G}_2(\hat\tau)-{\cal G}_2(x)\right|=0,
\label{oicfhw}\end{equation}
which can be also seen as an equation for
$x$, so that $v_0={\cal G}_2(x_0)$, $x_0=x_0(\hat\tau)$.

The second point concerns the parametrization of quantum vacua.
In this context we stress that since
$\hat \tau(u)$ is a univalent function,
that is the equality $\hat\tau(u_1)=\hat\tau(u_2)$ implies 
$u_1=u_2$, it follows that
the moduli space of quantum vacua can be equivalently identified with
the $\hat\tau$-space (or its fundamental domain $D_2$).
Therefore there is the correspondence
$$
u-moduli \; space \Longleftrightarrow
\hat\tau-moduli\; space.
$$
In the following we will use the subscript $\hat\tau$ to emphasize
the $\hat\tau$-parametrization of the vacuum states.
The above remarks suggest writing Eq.(\ref{uncertainty1}) in the form
(here $\hat \tau_0\equiv x_0$) 
\begin{equation}
\hbar |\langle {\rm tr}\,\phi^2\rangle_{\hat \tau}-
\langle {\rm tr}\,\phi^2\rangle_{\hat \tau_0}|
\leq 
{\pi}{\rm Im}\,\left(\langle\phi\rangle_{\hat \tau}
\overline{\langle\phi_D\rangle}_{\hat \tau}\right)\leq
2\hbar |\langle {\rm tr}\,\phi^2\rangle_{\hat \tau}-
\langle {\rm tr}\,\phi^2\rangle_{\hat \tau_0}|.
\label{aauncertainty1}\end{equation}
The fact that $\tau=\partial_a^2 {\cal F}$ is dimensionless implies that
$(2-a\partial_a){\cal F}=\Lambda\partial_\Lambda{\cal F}$, so that
Eq.(\ref{81}) is equivalent to
$\Lambda \partial_\Lambda {\cal F}=
-8\pi i  b_1\langle {\rm tr}\,\phi^2\rangle_{\hat \tau}$ where, as 
stressed in \cite{STY}\cite{EY}, 
$b_1=1/4\pi^2$ is the one-loop coefficient of the beta function. 
Thus Eq.(\ref{aauncertainty1}) implies
inequalities involving $\Lambda \partial_\Lambda {\cal F}$.
In \cite{STY} it has been suggested 
that the relation 
$\Lambda \partial_\Lambda {\cal F}=
-8\pi i  b_1\langle {\rm tr}\,\phi^2\rangle_{\hat \tau}$
should be understood in terms
of Renormalization Group ideas
(see \cite{STY}\cite{EY} for 
relevant generalizations of this formula and \cite{KKLMV}
for other interesting consequences).
In a forthcoming paper  \cite{BoMa} we will argue that
Eq.(\ref{aauncertainty1}) is related to the beta function 
$\Lambda\partial_\Lambda\tau$ whose structure can be investigated in the 
framework of the relation (\ref{81}) and Eq.(\ref{gdtf}).
We note that ${\cal C}$ can be seen as the curve which 
separates the local and asymptotic regions. Once the nonperturbative 
beta function is constructed the ray of convergence of the local
expansions
should be related to the structure of the $A$-domain and its boundary
${\cal C}$. In particular,  Eq.(\ref{aauncertainty1}) should also play a
a role in investigating Borel summability.

There is another aspect which should be mentioned
in discussing Eq.(\ref{uncertainty1}).
In ordinary Quantum Mechanics one has 
$\sqrt{\langle x^2\rangle_\psi-
\langle x\rangle_\psi^2}
\sqrt{\langle p^2\rangle_\psi-
\langle p\rangle_\psi^2}\ge \hbar/2$ (note that the square roots may be 
seen as Euclidean distances). Fields $\phi$ and $\phi_D$
play the role of $x$ and $p$ respectively and ${\cal F}$ is the
analog of the action. Similarly, one should consider the correspondence
$\langle x^2\rangle_\psi\to \langle {\rm tr}\,\phi^2\rangle_{\hat 
\tau}$ and
investigate whether 
$|\langle {\rm tr}\,\phi^2\rangle_{\hat \tau}-
\langle {\rm tr}\,\phi^2\rangle_{\hat \tau_0}|\sim 
|\langle {\rm tr}\,\phi^2\rangle_{\hat \tau}-
\langle \widehat\Theta\rangle_{\hat \tau}|$, with
$\widehat\Theta$ some field operator
(recall that by (\ref{oicfhw}) $\hat\tau_0$ is $\hat\tau$-dependent).
 To have a 
deeper analogy with the uncertainty relation 
one should identify $\langle {\rm 
tr}\,\phi_D^2\rangle_{\hat\tau}$
and investigate the structure of the relation between
$\langle {\rm tr}\,\phi^2\rangle_{\hat \tau_0}$,
$\langle {\rm 
tr}\,\phi^2\rangle_{\hat\tau}$,
$\langle\phi\rangle_{\hat\tau_0}$, $\langle\phi\rangle_{\hat\tau}$ 
and their duals.
The asymptotic behavior 
$\langle {\rm 
tr}\,\phi^2\rangle_{\hat\tau}\sim 
\langle\phi\rangle_{\hat\tau}^2/2$ 
suggests that
a similar relation should exist.
In this context it is interesting to note that 
$\langle\phi\rangle_{\hat\tau_0}^2$ 
appears in evaluating the Euclidean distance
$|\langle {\rm tr}\,\phi^2\rangle_{\hat \tau}-
\langle {\rm tr}\,\phi^2\rangle_{\hat \tau_0}|$. In particular,
deriving Eq.(\ref{81}) with respect to $\hat\tau$
we have
\begin{equation}
\partial_{\hat\tau}
\langle {\rm tr}\,\phi^2\rangle_{\hat \tau}
={1\over 8\pi i b_1} \langle \phi\rangle_{\hat\tau}^2,
\label{05}\end{equation}
so that Eq.(\ref{oicfhw}) becomes
\begin{equation}
a^2(v) \left(\overline{{\cal G}_2(\hat\tau)}-
\overline{{\cal G}_2(x)}\right)=
\overline{{a^2}(v)}\left({\cal G}_2(\hat\tau)-{\cal G}_2(x)\right).
\label{iuweh}\end{equation}

Finding the solution of Eq.(\ref{iuweh}) is an interesting open problem 
which should be possible to solve once the solution of (\ref{4})
is known.
As we said, solving Eq.(\ref{4}) is equivalent to find
the inverse of the map $\hat \tau={\cal H}(\tau)$ whose explicit 
expression is given by (\ref{14})(\ref{saouihkj}).
All these aspects show that the uncertainty relations (\ref{uncertainty1})
are described by the Schwarzian equation (\ref{4}).

Another aspect concerns the geometrical origin of the lower and upper 
bounds in (\ref{uncertainty1}). These seem to be a consequence of the good 
infrared and ultraviolet properties of negatively curved spaces.
These aspects have been investigated in a different context by 
G. Callan and F. Wilczek \cite{CallanWilczek}. Here we have 
seen that the Koebe 1/4-theorem
and Schwarz's lemma applied to hyperbolic geometry explain
the origin of such geometrical regulators. We also observe that the 
role of the Koebe 
1/4-theorem in hyperbolic geometry
seems related to the crucial mechanism which arises in 
the compactification of moduli spaces
of Riemann surfaces. In 
particular, the fact that in the
Deligne-Knudsen-Mumford (DKM)
compactification of moduli spaces of punctured
Riemann spheres (configuration space of anyons)
``punctures never collide'' can be seen as a request
for the Weyl-Petersson Hamiltonian to remain self-adjoint at the DKM
boundary. This implies
the mass-gap and the exclusion principle for anyons (punctures) \cite{mm3}.

There is another physical application of our results. Namely 
$a$ is a nowhere vanishing function of $u$ so that one should investigate
the role of the lower bound for $|a|$. This is relevant
in order to recover the structure of the $a$-moduli space and the 
properties
of $M_{n_mn_e}$ (such as the structure of the mass-gap). 
Similar arguments should be also useful to better 
understand some aspects concerning confinement.
In this context we recall that according to Nehari
theorem \cite{Nehari2}
a sufficient condition for the 
univalence of a function $g$ defined on $\Delta$ is
\begin{equation}
(1-|z|^2)^{2}|\{g,z\}|\le 2,
\label{38}\end{equation}
whereas the necessary condition is
\begin{equation}
(1-|z|^2)^{2}|\{g,z\}|\le 6.
\label{39}\end{equation}
It can be shown that the constant $2$ in (\ref{38})
cannot be replaced by any larger 
one. An interesting question
is to
find the sharp inequality for the case at hand.

\vspace{0.5cm}

In conclusion we note that our investigation is related with the theory
of quasidisks. They 
have interesting structures. For example a generic quasidisk has a
fractal boundary \cite{Bowen}.
Quasidisks also appear in some 
nonperturbative aspects of string theory \cite{Pekonen}.

\vspace{0.8cm}

 It is a pleasure to thank G. Bonelli, P.A. Marchetti and
M. Tonin for useful discussions.

\end{document}